\documentclass[twocolumn]{aastex631}

\usepackage{lineno}

\usepackage{float}

\usepackage{graphicx}

\usepackage{natbib}
\bibpunct{(}{)}{;}{a}{}{,}

\usepackage{booktabs}
\usepackage{makecell}
\usepackage{color, colortbl}

\usepackage[normalem]{ulem}
\usepackage{xcolor}

\usepackage{csquotes}

\begin{document}

\title{Electron scale magnetic holes generation driven by Whistler-to-Bernstein mode conversion in fully kinetic plasma turbulence}

\author[0000-0001-9180-5231]{Joaquín Espinoza-Troni}
\affiliation{Departmento de Física, Facultad de Ciencias,
Universidad de Chile, Santiago, Chile}

\author[0000-0001-7233-2555]{Giuseppe Arrò}
\affiliation{Los Alamos National Laboratory, Los Alamos, NM 87545, USA}

\author[0000-0002-7085-658X]{Felipe A Asenjo}
\affiliation{Facultad de Ingenier\'ia y Ciencias,
Universidad Adolfo Ib\'a\~nez, Santiago, Chile.}

\author[0000-0002-9161-0888]{Pablo S Moya}
\affiliation{Departmento de Física, Facultad de Ciencias,
Universidad de Chile, Santiago, Chile}



\begin{abstract}
Magnetic holes (MHs) are coherent structures characterized by a strong and localized magnetic field amplitude dip, commonly observed in the heliosphere. These structures come in different sizes, from magnetohydrodynamic to kinetic scales. Subion scale MHs are usually sustained by an electron current vortex and exhibit a strong electron temperature anisotropy, with higher temperatures perpendicular to the background magnetic field. Magnetospheric Multiscale observations (MMS) have revealed electron scale MHs to be ubiquitous in the turbulent Earth's magnetosheath and the solar wind, potentially playing an important role in the energy cascade and dissipation. Despite abundant observations, the origin of electron scale MHs is still unclear and debated. In this work, we use fully kinetic simulations to investigate the role of plasma turbulence in generating electron scale MHs. We find that the turbulence spontaneously produces electron scale MHs via the following mechanism: first, large-scale turbulent velocity shears produce regions with high electron temperature anisotropy; these localized regions become unstable, generating oblique electron scale whistler waves; as they propagate over the inhomogeneous turbulent background, whistler fluctuations develop an electrostatic component, turning into Bernstein-like modes; the strong electrostatic fluctuations produce current filaments that merge into an electron scale current vortex; the resulting electron vortex locally reduces the magnetic field amplitude, finally evolving into an electron scale MH. We show that MHs generated by this mechanism have properties consistent with MMS and nontrivial kinetic features with a "mushroom" shaped electron velocity distribution function. Our results have potential implications for understanding the formation and occurrence of electron scale MHs in astrophysical turbulent and space environments, such as the Earth's magnetosheath and the solar wind.
\end{abstract}



\section{Introduction}

Magnetic holes (MHs) are coherent structures usually observed in turbulent plasma environments and are characterized by a sharp decrease in the magnitude of the magnetic field. The structures were first detected by \citet{Turner_etal_1977} in the Solar Wind (SW) at 1 AU and defined as isolated depressions in the interplanetary magnetic field (IMF) intensity, which is otherwise nearly constant, distinguishing them from random fluctuations. Since then, countless observations of MHs in the solar wind have been reported \citep{Winterhalter_etal_2000, Stevens_etal_2007, Karlsson_etal_2021, Wang_etal_2021, Yu_etal_2021}, and observed in the IMF from $0.3$ AU to $17$ AU, using data from Helios 1,2 and Voyager 2 \citep{Sperveslage_etal_2000}. It has been reported that MHs are abundant not only in the SW but also in other space environments, as in the terrestrial and planetary magnetosheaths \citep{Johnson_etal_1997, Soucek_etal_2008, Volwerk_etal_2008, Huang_etal_2017b, Yao_etal_2017, Zhong_etal_2019, Goodrich_etal_2021, Karlsson_etal_2021, Chen_etal_2022, Chen_etal_2023, Yao_etal_2023,Xu_etal_2024, Wang_etal_2024a,Wang_etal_2024b}, magnetospheric cusps \citep{Shi_etal_2009} and around comets \citep{Russell_etal_1987}.  

MHs can exist on different scales. The largest MHs can extend up to hundreds of ion gyroradii $\rho_i$ at fluid magnetohydrodynamic scales \citep{Stevens_etal_2007}. In the following, we will refer to these structures as \enquote{large scale MHs}. In contrast, sub-ion scale MHs have sizes ranging from $\rho_i$ to a few electron gyroradii $\rho_e$ \citep{Yao_etal_2017}. Large-scale MHs are characterized by a temperature anisotropy in ions with $T_{i\perp} > T_{i\parallel}$, where $T_{i\perp}$ and $T_{i\parallel}$ are the ion temperature components perpendicular and parallel to the background magnetic field, respectively. They also present an increase in density and temperature, which tend to balance the magnetic pressure decreases due to the magnetic field depression. These structures have attracted great interest because they potentially play a role in regulating the temperature anisotropy of turbulent plasmas such as the magnetosheath and the SW \citep{Yu_etal_2021}. It has also been shown that they can generate waves due to their temperature anisotropy \citep{Ahmadi_etal_2018, Shahid_etal_2024}. 

In recent years, sub-ion scale MHs have been a source of great interest due to the Magnetospheric Multiscale (MMS) mission, which has opened the possibility to obtain data at kinetic scales with great precision \citep{Burch_etal_2015}. The MMS mission has made it possible to detect many small MHs down to scales of a few electron gyroradii. These structures have been observed in the solar wind \citep{Wang_etal_2020} and the terrestrial plasma sheet \citep{Sundberg_etal_2015,Goodrich_etal_2016}, but their rate of occurrence is greater in the Earth's magnetosheath \citep{Shutao_Yao_etal_2021}. A recently published review discusses the latest observations and studies regarding these structures observed by MMS in the Earth's magnetosheath \citep{Shi_etal_2024}. Sub-ion scale MHs are characterized by an increase in the perpendicular electron temperature producing electron temperature anisotropy, caused by trapped populations of hot electrons with high-pitch angles. As in the case of large-scale MHs, these trapped particles increase the density inside the MH and generate a pressure gradient that balances the decrease of the magnetic field pressure \citep{Liu_etal_2019}. Nevertheless, these structures present no correlation with the ions due to their small scale, and usually, they have an ion temperature $T_i$ larger than the electron temperature $T_e$ \citep{Liu_etal_2021}. 

Sub-ion scale MHs are generally found in turbulent environments, and it is thought that they may play a role in their dynamics \citep{Shi_etal_2024}. Turbulence in collisionless magnetized plasmas is critical in transferring energy from large MHD scales to small kinetic scales where energy is dissipated. This energy cascade has been widely explored, especially in the heliosphere for its potential role in heating the solar corona and accelerating the solar wind \citep{Sahraoui_etal_2020}, but also in the magnetosheath for its importance in driving energy transfer processes and particles into the inner magnetosphere through the magnetopause \citep{Huang_etal_2014, Narita_etal_2016, Rakhmanova_etal_2021, Arro_etal_2022, Gallo_Moya_2023}. Unlike neutral fluids, where the dissipation mechanism on small scales is due to the viscosity resulting from molecular collisions, in non-collisional plasmas it is due to collisionless kinetic processes such as wave-particle interactions \citep{Sahraoui_etal_2020}. In this context, sub-ion scale MHs produced in the magnetosheath have aroused great interest since it is thought that they might participate in small-scale energy dissipation processes. Their role in the generation of waves \citep{Huang_etal_2018, Yao_etal_2019}, in the acceleration of electrons, and plasma heating \citep{Shi_etal_2024} has been studied extensively. Sub-ion scale MHs have also been linked to magnetic reconnection processes \citep{Zhong_etal_2019, Li_and_Zhang_2023}, and inside them are usually observed non-Maxwellian distributions that could affect the dynamics of the plasma \citep{Arro_etal_2023}. It is worth mentioning that there is still no conclusive evidence regarding the impact and the role of sub-ion scales MHs on energy dissipation, an issue that is still debated and under investigation.

Various mechanisms have been proposed for the formation of MHs, which vary depending on their scales. To explain the formation of large-scale MHs, some of these mechanisms include the ponderomotive force due to the propagation of phase steepened Alfvén waves in the solar wind \citep{Tsurutani_etal_2002_b, Dasgupta_etal_2003, Tsurutani_etal_2005}, and the non-linear evolution of the mirror instability \citep{Zhang_etal_2008, Ahmadi_etal_2018}. MHD solitons have also been proposed to describe their final fully developed stage \citep{Baumgartel_etal_2003}. Satellite observations have shown that the structures produced by the nonlinear mirror instability share features consistent with large-scale MHs, in addition to the fact that the threshold condition of mirror instability requires ion temperature anisotropy with $T_{i\perp} > T_{i\parallel}$ to develop \citep{Kivelson_etal_1996, Baumgartel_etal_2003, Kuznetov_etal_2007, Soucek_etal_2008}. Therefore, the nonlinear evolution of mirror instability has attracted great attention as a possible mechanism for generating large-scale MHs. However, these nonlinear structures produced by the mirror instability can manifest as peaks or depressions of the magnetic field. Numerical simulations have shown that localized dips in the magnetic field are not a typical feature of saturated states of mirror instability, that instead tend to produce magnetic peaks; in contrast, isolated MHs form mainly in regions where the plasma is mirror stable with low values of the ion plasma beta  \citep{Baumgartel_etal_2003, Califano_etal_2008, Ahmadi_etal_2017}, which also agrees with statistical analysis of satellite observations \citep{Soucek_etal_2008}. \citet{Shoji_etal_2012}, using hybrid simulations, analyzed the relation between the mirror instability and magnetic peaks and dips observed in the magnetosheath. The authors have shown that the nonlinear stage of the mirror instability produces MHs only in 2D geometries with low ion-beta conditions, a condition hardly met in the magnetosheath. Therefore, mirror instability as a mechanism for large-scale MHs generation requires very specific plasma conditions, and its role in this context is still debated, with no clear evidence of it being the cause of the formation of large-scale MHs in the magnetosheath.

For the formation of MHs at sub-ion scales, other mechanisms different than those mentioned for large-scale MHs have been proposed. Some of them include the electron mirror/field-swelling instability \citep{Basu_and_Coppi_1982, Marchenko_etal_1988, Gary_etal_2006, Pokhotelov_etal_2013, Hellinger_Stverak_2018, Liu_etal_2021}, but this requires the condition that $T_e > T_i$ for the formation of sub-ion scale MHs, which is usually not satisfied in observations of small-scale MHs \citep{Liu_etal_2021}. However, in \citet{Yao_2019} it has been assessed that even if $T_e > T_i$, by considering the large-scale differences between electrons and ions, at ionic scales ions can be treated as a cold background and the electron mirror instability can be excited, producing sub-ion scale MHs. MHs have also been analytically described using the propagation of electron magnetohydrodynamic (EMHD) solitons in the plasma sheet \citep{Ji_etal_2014, Li_etal_2016}. This is consistent with some observations, even if this description is not suitable for non-propagating MHs in the plasma rest frame, as those observed in some numerical simulations \citet{Haynes_etal_2015}, and other MHs detected by satellite data \citep{Liu_etal_2019}. Therefore, there is no conclusive evidence that the instabilities mentioned above are capable of producing MHs at both sub-ion and large scales.

In addition to being studied via observations in the magnetosheath, sub-ion scale MHs have been analyzed via 2D and 3D Particle-in-cell (PIC) simulations of freely decaying turbulence \citep{Haynes_etal_2015,Roytershteyn_etal_2015, Arro_etal_2023}, and have been studied theoretically using self-consistent kinetic models based and tested on observational data \citep{Li_etal_2020}. Turbulence is known to produce long-lived coherent structures \citep{servidio2012local,karimabadi2013coherent,lion2016coherent,perrone2016compressive,arro2024spatio}, so it represents one of the best candidates to drive the formation of MHs. In \citet{Haynes_etal_2015}, it is shown that magnetic field depressions are non-propagating structures associated with an azimuthal diamagnetic current provided by a population of trapped electrons following petal-like orbits. Because of this, such structures have been dubbed "electron vortex magnetic holes" (EVMHs). The properties of EVMHs are consistent with numerous observations \citep{Shi_etal_2024}. For ion scales, a similar structure can occur, and it has been observed that an ion ring-like current can sustain an ion vortex magnetic hole (IVMH) with a self-consistent structure, which makes it stable for a long time \citep{Wang_etal_2021, Yao_etal_2023}. In the work of \citet{Haynes_etal_2015} the turbulence is initialized with fluctuations whose wavelengths are close to the size of the resulting sub-ion scale MHs, implying that this structure is likely a result of the initial perturbations. For this reason, previous numerical works, despite describing the structure of EVMHs, do not address the specific mechanism by which the large-scale turbulent dynamics that are found in the magnetosheath evolve and produce the plasma conditions necessary for the generation of sub-ion scale MHs. 

Recent numerical studies have shown evidence for the formation of MHs due to self-consistent turbulent mechanisms at both large and sub-ion scales. In \citet{Arro_2024}, a possible mechanism for the formation of large-scale MHs is analyzed using a 2D hybrid simulation initialized with solar wind parameters. In this study, MHs emerge from initial magnetic field perturbations by trapping hot ions with large pitch angles, as in a magnetic mirror, making the structure stable for hundreds of ion gyroperiods. In the case of sub-ion scale MHs, \citet{Arro_etal_2023} discuss a possible turbulent-driven mechanism for generating EVMHs using a fully kinetic 2D simulation of freely decaying turbulence initialized with magnetosheath parameters. In this study, turbulence is initialized with fluctuations whose wavelengths span more than $10$ ion inertial lengths $d_i$, much larger than the size of EVMHs. The development of electron velocity shears at large scales due to the evolution of turbulence leads to the electron Kelvin-Helhomtz instability, which breaks the electron velocity shears into small-scale electron vortices. Then, the electron current carried by these vortices reduces the local magnetic field intensity, producing EVMHs, whose size is of the order of $1\,d_i$. Hence, the aforementioned numerical study discusses a possible mechanism that contributes to understanding how large-scale turbulent fluctuations can dynamically set up the conditions for the formation of sub-ion scale MHs. 

In this work, we discuss another possible turbulent-driven mechanism for the formation of sub-ion scale MHs. We use the same fully kinetic simulation used in \citet{Arro_etal_2023} to study the role of turbulence in producing EVMHs at electron scales, with a size of a few electron inertial lengths $d_e$. The EVMHs that we will discuss are even smaller than the ones analyzed in \citet{Arro_etal_2023}, which have near-ion scales with sizes close to an ion inertial length. As we are mainly interested in the EVMHs observed in the magnetosheath, we perform the simulations with parameters consistent with those typically observed in this environment. Also, we initialize the turbulence with large-scale fluctuations that span several ion inertial lengths $d_i$. In this way, since the injection scale is very far from electron scales, we ensure that the EVMHs are a consequence of the turbulence energy cascade instead of being a direct consequence of the initial perturbations that drive the turbulence. We find that electron scale MHs can be generated by turbulence via the following mechanism: first, the shearing motion induced by large-scale velocity fluctuations produces localized regions with high electron temperature anisotropy, providing free energy for the development of the oblique whistler-cyclotron instability; then, due to the non-linear interaction of whistler waves with the turbulent and inhomogeneous background, these modes convert into oblique Bernstein waves; the strong electrostatic fluctuations associated with Bernstein modes, induce electron drift currents in the form of a train of vortices; these vortices finally merge into a single larger vortex, reducing the local magnetic field magnitude, ultimately evolving into an EVMH. This process serves as a possible explanation for the formation of electron-scale MHs, driven by the evolution of large-scale fluctuations in the turbulence cascade. The observed EVMHs have nontrivial kinetic properties with a "mushroom" shaped electron velocity distribution function (EVDF) and have features that are consistent with MMS observations.

\section{Simulation Setup}

We study the formation of electron scale MHs in a fully kinetic simulation of freely decaying plasma turbulence, which is performed using the semi-implicit energy-conserving PIC code ECsim \citep{Markidis_and_Lapenta_2011, Lapenta_2017, Lapenta_2023}. A 2D square period box represents the simulation domain whose size is $L = 64 d_i$ and is sampled by a uniform mesh containing $2048^2$ points. We consider an ion-electron plasma with a reduced ion-to-electron mass ratio of $m_i/m_e = 100$. Both species are initialized from a Maxwellian distribution, with $5000$ particles per cell per species and uniform density. The initial conditions to set up the plasma simulation are chosen to reproduce similar conditions as those observed by satellite measurements in the terrestrial magnetosheath \citep{Phan_etal_2018, Stawarz_etal_2019, Bandyiopadhyay_etal_2020}. The plasma is initially quasi-neutral with uniform and isotropic temperature, with plasma beta equal to $\beta_i = 8$ for ions and $\beta_e=2$ for electrons. The electron inertial length is $d_e = 0.1 d_i$, while the ion and electron gyroradii are initially equal to $\rho_i=\sqrt{\beta_i}d_i \simeq 2.83 d_i$, and $\rho_e=\sqrt{\beta_e}d_e \simeq 1.41 d_e$, respectively. Also, we consider initially an out-of-plane magnetic field configuration $B_0 \hat{z}$ (with $\hat{z}$ being the unit vector in the out-of-plane direction). Turbulence is triggered by random phase isotropic magnetic field and velocity perturbations. The wavenumbers $k$ of the initial perturbations fall in the range $1 \leqslant k/k_0 \leqslant 4$ (where $k_0=2\pi/L$), which corresponds to an injection scale of about $\lambda_{inj} = 16 d_i$, much larger than $d_e$. The root mean square (rms) amplitude of magnetic field fluctuations $\delta B$ is $\delta B_{rms}/B_0 = 0.9$ while the rms amplitude of ion and electron fluid velocity fluctuations are $\delta u$ is $\delta u_{rms}/c_A = 3.6$ (with $c_A$ being the initial Alfvén speed based on the initial guide field $B_0$). In addition, the ratio between the electron cyclotron frequency $\Omega_e$ and the ion cyclotron frequency $\Omega_i$ is $\Omega_e/\Omega_i = m_i/m_e = 100$, while the ratio between the plasma frequency and the cyclotron frequency is $\omega_{pi}/\Omega_i = 100$ for ions and $\omega_{pe}/\Omega_e = 10$ for electrons. The time step used in the simulation is equal to $\delta t = 0.05\Omega_e^{-1}$. We have solved the simulation up to time $t=500\Omega_e^{-1}$ when the turbulence is fully developed. Additional information about this simulation and its turbulent properties are provided in \citet{Arro_etal_2022}.

\section{Results}

\subsection{Fully developed Electron Scale Magnetic Hole}

\begin{figure*}[ht]
\centering
\includegraphics[width=0.92\linewidth]{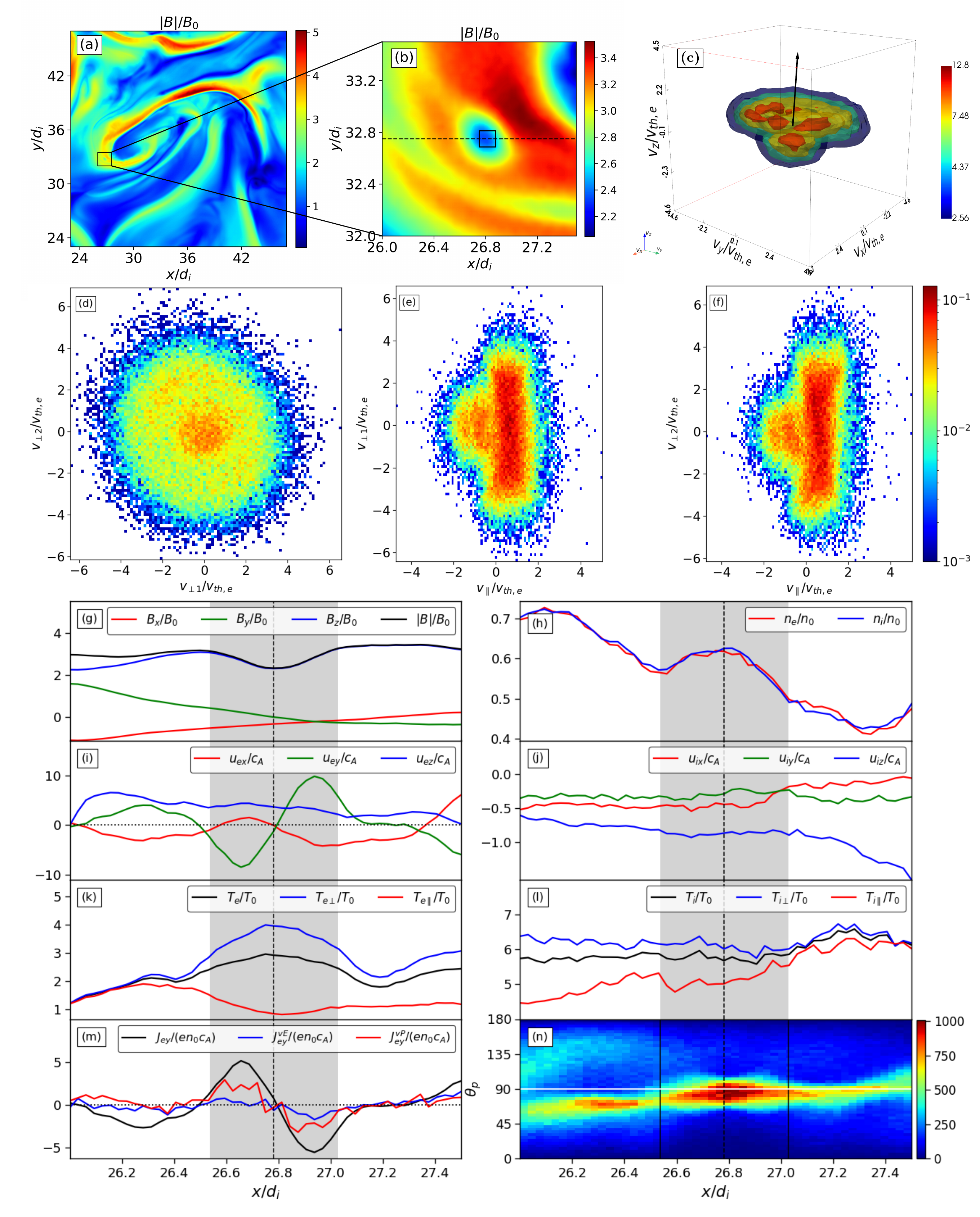}
\caption{Panels (a)-(b): shaded isocontours of the magnetic field magnitude over a certain region ($x/d_i \in [23,47]$ and $y/d_i \in [23,47]$) of the simulation box at $t = 500\Omega_e^{-1}$, with a zoom showing an electron scale MH. Panel (c): 3D isosurfaces of the EVDF taken in the region marked by the black square in panel (b), with the black arrow indicating the direction of the mean local magnetic field. Panels (d)-(f): 2D projections of the EVDF in the orthogonal system of coordinates aligned with the mean local magnetic field with $\hat{\mathbf{x}}_{1\perp} = \hat{\mathbf{z}}\times \bar{\mathbf{B}}/\sqrt{\langle \bar{B}_x\rangle^2+\langle \bar{B}_y\rangle^2}$, $\hat{\mathbf{x}}_{2\perp} = \bar{\mathbf{B}} \times \hat{\mathbf{x}}_{1\perp} /\left(| \bar{\mathbf{B}} |\sqrt{\langle \bar{B}_x\rangle^2+\langle \bar{B}_y\rangle^2}\right)$ and $\hat{\mathbf{x}}_\parallel = \bar{\mathbf{B}}/|\bar{\mathbf{B}}|$.} Panels (g)-(m): magnetic field, electron and ion number densities, electron and ion fluid velocities, electron and ion temperatures, and electron drift currents, over a 1D cut crossing the MH, indicated by the black dashed line in panel (b). In these panels, the gray shaded area highlights a region of width $ 0.5 d_i$ around the MH, whose center is indicated by the black vertical dashed line. Panel (n): Electron pitch angle spectrum (in degrees) over a stripe-shaped region of width $0.125 d_i$ around the black shaded line of panel (b), with the color bar representing the number of particles.
\label{fig:MH_characteristics}
\end{figure*}

\begin{figure*}[ht]
\centering
\includegraphics[width=\linewidth]{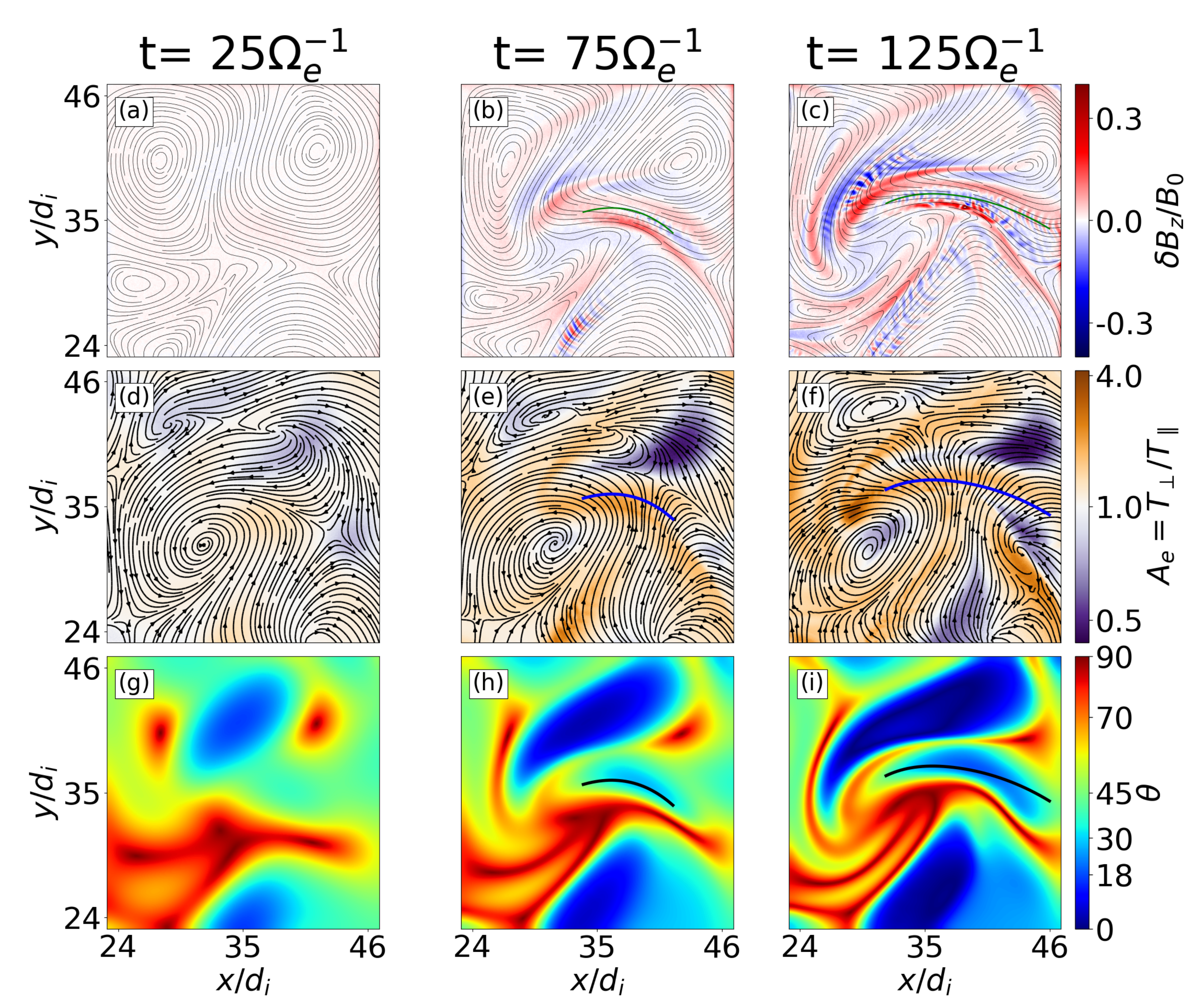}
    \caption{Panels (a)-(c): shaded isocontours of the high pass filter of the out-of-plane component of the magnetic field $\delta B_z$ with the black streamlines representing the in-plane magnetic field. Panels (d)-(f): shaded isocontours of the electron anisotropy $A_e$ with the black streamlines representing the in-plane electron fluid velocity. Panels (g)-(i): shaded isocontours of the angle $\theta$ between the low-pass filtered magnetic field and the plane, i.e $\theta = \arctan\{\langle B_z\rangle/\sqrt{\langle B_x\rangle^2+\langle B_y\rangle ^2}\}$. The panels are divided into three columns showing the time sequence in the following times $t=25\Omega_e^{-1}$, $t=75\Omega_e^{-1}$, and $t=125\Omega_e^{-1}$. The lines marked on times $t=75\Omega_e^{-1}$ and $t=125\Omega_e^{-1}$ represent the 1D cut crossings used to compute the quantities for the linear solver analysis and show the wave characteristics (see below).}
\label{fig:Anisotropy_instability}
\end{figure*}

\begin{figure*}[ht]
\centering
\includegraphics[width=\linewidth]{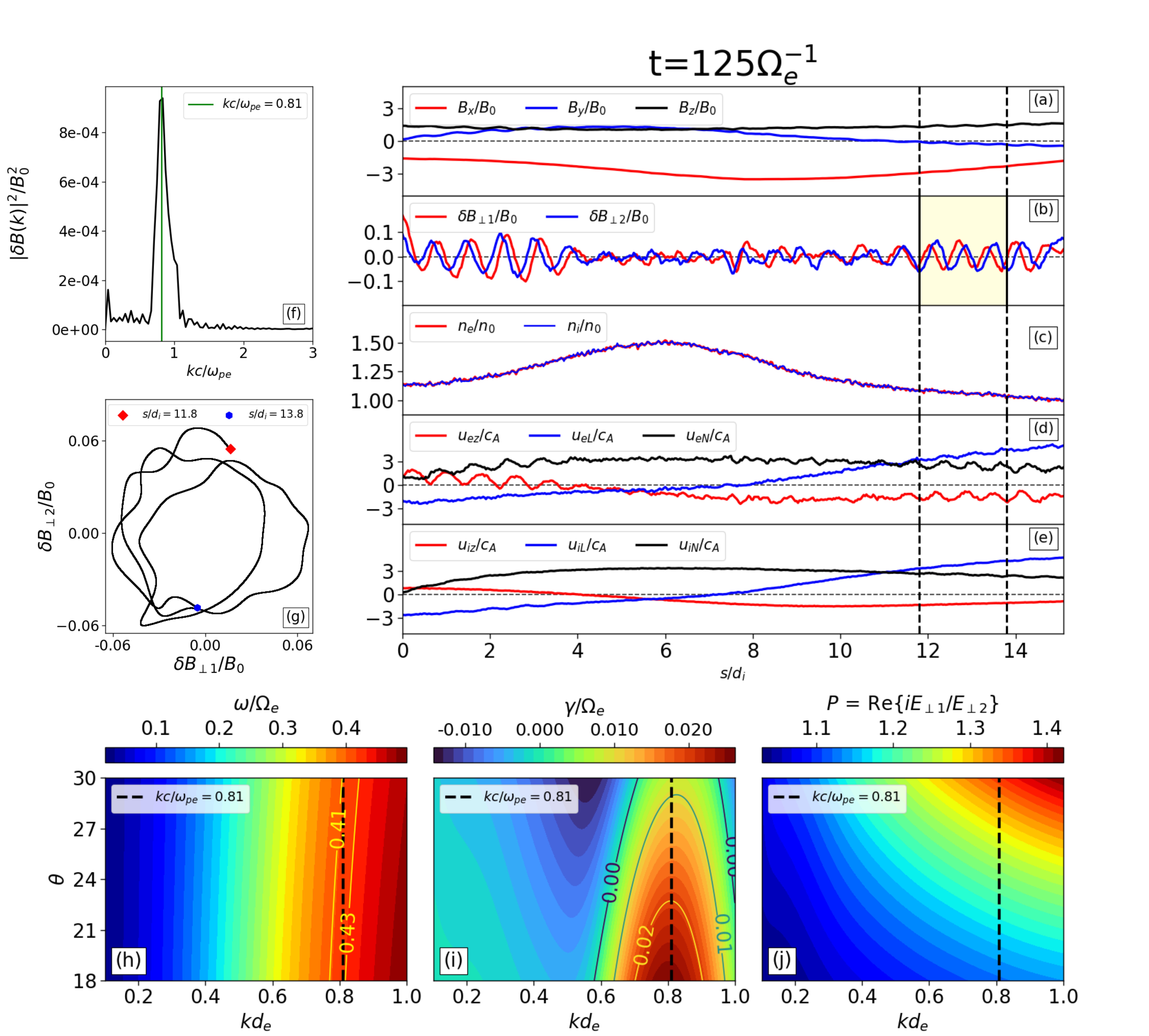}
    \caption{Panels (a)-(e): components of the total magnetic field, components of the high pass filtered magnetic field in the orthogonal system of coordinates aligned with the low pass filtered magnetic field, with $\hat{\mathbf{x}}_{1\perp} = \hat{\mathbf{z}}\times \langle \mathbf{B}\rangle/\sqrt{\langle B_x\rangle^2+\langle B_y\rangle^2}$ and $\hat{\mathbf{x}}_{2\perp} = \langle \mathbf{B}\rangle \times \hat{\mathbf{x}}_{1\perp} /\left(|\langle \mathbf{B}\rangle |\sqrt{\langle B_x\rangle^2+\langle B_y\rangle^2}\right)$, electron and ion density, electron and ion components of the fluid velocity in LNZ coordinates over a 1D cut crossing marked by a line in the panels of the third column of Figure \ref{fig:Anisotropy_instability} for $t=125\Omega_e^{-1}$. Panel (f): square magnitude of the FFT of the high-pass filtered magnetic field shown on panel (b) as a function of the wavenumber normalized to the electron inertial length. Panel (g): hodogram using the high-pass filtered magnetic field components over the region highlighted in yellow on panel (b). Panels (h)-(j): shaded isocontours of the real and imaginary components of the frequency normalized to the electron gyrofrequency, and the electric field polarization as a function of the wavenumber $k d_e$ and the angle between the wave direction and the magnetic field $\theta$ as it is predicted by the linear solver using the data of table \ref{tab:LinearSolverData}. The vertical black shaded line marks the wavenumber predicted by the FFT on panel (f).}
\label{fig:Whistler_characteristics}
\end{figure*}

In this subsection, we study the properties of an electron scale MH that we observe when turbulence is fully developed. Panels (a) and (b) of Figure \ref{fig:MH_characteristics} show the shaded isocontours of the magnetic field magnitude $|B|$ normalized to the initial magnitude of the magnetic field $B_0$ with a progressive zoom on the MH. These panels visualize the MH in the 2D turbulence simulation box for $t=500\Omega_e^{-1}$.
Here, we can observe the electron scale MH as a stable, coherent structure coexisting with the completely evolved plasma turbulence. In panel (b) of Figure \ref{fig:MH_characteristics}, we can identify the MH as a depression of the magnetic field magnitude of $\sim 30\%$ with a length-scale of $\sim 5 d_e$, which is at the same scale as the MH observed in magnetosheath satellite measurements reported in \citet{Yao_etal_2017} and the simulations in \citet{Haynes_etal_2015}. In Figure \ref{fig:MH_characteristics}.(c) we represent 3D isosurfaces of the EVDF calculated inside the MH, in the region marked by the black square in panel (b) of the same figure, with velocities normalized to the initial electron thermal velocity $v_{th,e}$. In this panel, the black arrow represents the direction of the average magnetic field over the sampled region. We can see that the EVDF has a "mushroom" shape, with two populations: a hot anisotropic ring that acts as the "hat" of the "mushroom" and a colder and more isotropic population that forms the "foot" of the "mushroom", both aligned to the parallel direction of the magnetic field, with higher temperatures in the direction perpendicular to the magnetic field. We call the "hat" population a ring because it is void inside, as we see in panel (c). The "hat" is not centered at $v=0$ and is displaced in the positive parallel direction with respect to the magnetic field, while the "foot" is displaced in the opposite direction. Panels (d)-(f) of Figure \ref{fig:MH_characteristics} show the 2D projections of the EVDF in the orthogonal system of coordinates aligned with the mean local magnetic field $\bar{\mathbf{B}}$. It is clear on panel (d) that the EVDF is isotropic in the plane perpendicular to the local mean magnetic field. Also, we can identify in panels (e) and (f) the cold and hot populations that form the EVDF, with the last one having a high electron temperature anisotropy. Panels (g)-(m) of Figure \ref{fig:MH_characteristics} show different plasma quantities along a horizontal cut crossing that passes through the electron scale MH, marked with a black dashed line in Figure \ref{fig:MH_characteristics}.(b). Panel (g) shows the different components of the magnetic field and its total magnitude. It is clear from this panel that there is a depletion in the magnetic field magnitude, with the magnetic field pointing in a direction almost perpendicular to the x-y plane, as the magnetic field magnitude is almost equal to the $B_z$ component. We have highlighted in all the panels (g)-(n) of figure \ref{fig:MH_characteristics} the region associated with this depletion, i.e. the MH, represented by the gray shaded area, with the position of the minimum of the magnetic field indicated by a vertical black dashed line. Figure \ref{fig:MH_characteristics}.(h) shows the electron and ion densities normalized to the initial density $n_0$. From this figure, we see that the plasma is quasi-neutral, and we can observe an enhancement of density inside the MH region. This is a typical MH feature that is consistent with observations of small-scale MHs \citep{Shi_etal_2024}. This increase in density is consistent with the idea that the MH is trapping particles. A possible mechanism that could be playing a role in this trapping is the one studied in \citet{Haynes_etal_2015} for EVMHs, where electrons are trapped via the $\nabla B$ drift associated with the magnetic field drop, which causes these particles to move in petal-shaped orbits. Due to the small scale of the structure, the ions are not coupled to it. Indeed, there is no correlation between ion quantities and the MH region. Nevertheless, ions are probably gathered into the MH due to the electrostatic potential produced by the concentration of electrons, maintaining quasi-neutrality. Panels (i) and (j) of Figure \ref{fig:MH_characteristics} show the components of the electron and ion fluid velocity $\mathbf{u}_e$ and $\mathbf{u}_i$, respectively, normalized to the initial alfvén velocity $c_A$. The ion velocity does not show any correlation with the magnetic field magnitude drop. On the contrary, the in-plane components of the electron velocity show a reversal across the MH, with $u_{e,y}$ (perpendicular to the cut) being the dominant component, indicating the presence of a ring current structure. Therefore, the MH is sustained mostly by the electron current, with almost no participation of the ions, whose velocity is 2 orders of magnitude less than the electron's velocity. Panels (k) and (l) of Figure \ref{fig:MH_characteristics} show the scalar temperature $T$, and the components of the temperature parallel $T_\parallel$ and perpendicular $T_\perp$ to the local magnetic field for electrons and ions respectively, normalized to the initial temperature $T_0$. We can see that inside the MH there is an enhancement in $T_{e\perp}$ and a reduction in $T_{e\parallel}$, with an increase in the total electron temperature. On the other hand, the behavior of the ion's temperature components does not show any variation correlated with the magnetic dip. Hence, there is an increase in the electron temperature anisotropy $A_e=T_{e\perp}/T_{e\parallel} > 1$ inside the electron scale MH, and $T_i > T_e$, consistent with typical satellite observations of sub-ion scale MHs \citep{Huang_etal_2017}.

Figure \ref{fig:MH_characteristics}.(m) compares the $y$ components of the electron current density $\mathbf{J}_e$, the electron $\textbf{E}\times\textbf{B}$ drift current $\mathbf{J}_e^{vE} = -en_e c \mathbf{E}\times\mathbf{B}/|\mathbf{B}|^2$, and the electron pressure drift current $\mathbf{J}_e^{vP} = -c(\nabla\cdot \mathbf{P}_e)\times \mathbf{B}/|\mathbf{B}|^2$. These quantities are normalized to $en_0 c_A$. As the density and temperature increase inside the MH, a pressure gradient tends to balance the electron current density to sustain the structure \citep{Liu_etal_2019}. Indeed, if we only consider the pressure and hall terms in the generalized Ohm law, it predicts that $J_e \approx J_e^{vE} + J_e^{vP}$. We can neglect the MHD term since we are looking at electron scales. The pressure gradient current sustains a significant part of the current, and although small, there is a contribution from the $\textbf{E}\times\textbf{B}$ drift current. Similar properties have also been observed in sub-ion scale MHs in satellite data \citep{Li_etal_2020}. Nevertheless, the sum of the pressure gradient current and the $\textbf{E}\times\textbf{B}$ drift current does not match exactly the total electron current density. This is possibly related to the fact that since the MH size is of the order of electron scales, inertial terms in the generalized Ohm's law and kinetic effects may contribute significantly to the total current, explaining the observed mismatch. 

In panel (n) of Figure \ref{fig:MH_characteristics}, we show the electron pitch angle distribution over a stripe-shaped region of width $0.125 d_i$ centered around the 1D cut crossing that passes through the MH in panel (b). To calculate the electron pitch angle, we subtracted the local electron fluid velocity $\mathbf{u}_e$ from the velocity of each electron $\mathbf{v}_e$. To obtain the local electron bulk velocity at each particle position, we interpolated the electron fluid velocity. Then, the pitch angle $\theta_p$ is given by $\theta_p = \cos^{-1}\left((\mathbf{v}_e-\mathbf{u}_e)\cdot \mathbf{B}/|\mathbf{B}||\mathbf{v}_e-\mathbf{u}_e| \right)$. From panel (n), we can see a trapped population of electrons inside the region occupied by the MH. This population is centered at a pitch angle of almost $90$ degrees. This shows that the MH hosts particles with high pitch angles, in agreement with what has been observed in simulations \citep{Haynes_etal_2015}, and in observational data \citep{Li_etal_2020}, for sub-ion scale MHs. Also, this high-pitch angle electron-trapped population causes the increase of electron anisotropy, which is usually observed in MHs. 

In the following sections, we will follow the MH formation over time, identifying the different steps and mechanisms involved in this turbulence-driven process. We will show how large-scale turbulent velocity shears induce the development of regions characterized by enhanced perpendicular electron temperature anisotropy. Then we will demonstrate that the electron temperature anisotropy makes the plasma unstable to the oblique whistler-cyclotron instability and leads to the generation of waves. Later, we will describe how these waves develop an electrostatic component with Bernstein-like properties. Finally, we will analyze how the formation of current filaments correlated with the electrostatic fluctuations gives rise to an electron vortex that produces the electron scale MH described above.

\subsection{Whistler wave generation} 

We will follow the electron scale MH formation from $t=25\Omega_e^{-1}$, when turbulence is not fully developed. To analyze the sub-ion scale magnetic field fluctuations that the turbulence dynamics produces we show in panels (a)-(c) of Figure \ref{fig:Anisotropy_instability} the high pass filtered out-of-plane component of the magnetic field $\delta B_z$ with the in-plane magnetic field streamlines represented by black lines for times $t\Omega_e \in \{25,75,125\}$. We used a high-pass filter performed with a spatial Gaussian filter with a length of $0.75 d_i$. This length is chosen to exclude large-scale fluctuations. Selecting fluctuations at scales of the order of those of the wave, we will analyze in the following. Hereafter, we will use this filtering method with the discussed length for all low and high pass-filtered quantities. We can observe in panel (a) that at time $t=25\Omega_e^{-1}$, there are almost no fluctuations in $\delta B_z$. As time progresses, we can see in panel (b) that at time $t=75\Omega_e^{-1}$, significant fluctuations in the $\delta B_z$ appear. Later, at time $t=125\Omega_e^{-1}$, the amplitude of fluctuations increases, reaching values of the order of $10^{-1} B_0$. We will focus on analyzing fluctuations sampled over the green 1D lines highlighted in the figures, as those fluctuations will later produce the MH. We also observe that the fluctuations look like a wave propagating in the direction parallel to the in-plane magnetic field. To understand the emergence of these fluctuations we show in panels (d)-(f) of Figure \ref{fig:Anisotropy_instability} a time-sequence of the shaded isocontours of the electron temperature anisotropy $A_e = T_\perp/T_\parallel$, with the streamlines of the electron fluid velocity represented by black lines. Another relevant quantity needed to discuss the evolution of these fluctuations is the angle $\theta = \arctan\left(B_z/\sqrt{B_x^2+B_y^2}\right)$ between the total magnetic field and the x-y plane, represented in panels (g)-(i) of Figure \ref{fig:Anisotropy_instability}. Since the fluctuation we are interested in propagates in the direction parallel to the in-plane magnetic field, our definition of $\theta$ is a good approximation for the angle between the background magnetic field and the direction of wave propagation. The green cuts of panels (a)-(c), which are selected to cross the fluctuations we want to analyze, are the same as the blue cuts in panels (d)-(f) and black cuts in panels (h)-(i). We can see in panel (e) for time $t=75\Omega_e^{-1}$, that in the same region where magnetic fluctuations appear, there is an electron velocity shear on the x-direction of the form $\sim \partial_x u_{ey}$. Furthermore, in this region, the electron temperature anisotropy is increasing. Also, we see in panel (h) that the out-of-plane electromagnetic fluctuation propagates with a relatively small angle with respect to the background magnetic field, ranging between $18$ and $30$ degrees. In the third column of Figure \ref{fig:Anisotropy_instability}, at $t=125\Omega_e^{-1}$, there is a stronger enhancement of the electron temperature anisotropy in correspondence of the electron velocity shear region, as seen in panel (f). This correlates with the larger amplitude of magnetic field fluctuations propagating in a direction almost parallel to the local magnetic field, as shown in panels (c) and (i). 

Considering the evolution described above, we interpret the development of the observed electromagnetic fluctuations in the following way. Due to the large-scale electron velocity shear indicated by the streamlines in panels (e)-(f) of Figure \ref{fig:Anisotropy_instability}, the electron temperature anisotropy increases locally. This is a well-studied mechanism for non-gyrotropic and gyrotropic temperature anisotropy generation in plasmas \citep{DelSarto_etal_2016}. The contribution of the shear in the pressure tensor can be calculated directly from the second moment of the Vlasov equation, as it has been done in \citet{DelSarto_and_Pegoraro_2017}, assuming a negligible heat flux contribution. Indeed, as described in \citet{Pucci_etal_2021}, for an initially isotropic plasma with pressure $P$ and no magnetic field, the equation of evolution for the pressure tensor $\mathbf{P}$ is $\partial \mathbf{P}/\partial t = -P\mathbf{S}$, with $S_{ij} = \left[\partial_j u_i + \partial_i u_j\right]$ being the velocity stress tensor. This equation shows the effect of the velocity shear on the pressure tensor, whose temporal derivative is proportional to the stress tensor. Also, as we will show later, the background magnetic field increases its magnitude mostly in the $-\hat{x}$ direction, due to the compression term $\partial_x u_x$. This is consistent with the EMHD version of the induction equation for the magnetic field $D\mathbf{B}/Dt = \mathbf{B}\cdot \nabla \mathbf{u}_e + \mathbf{B}\nabla\cdot \mathbf{u}_e $, with $D/DT = \partial_t + \mathbf{u}_e \cdot \nabla$ \citep{Lyutikov2013}, which is suited to describe the magnetic field dynamics at near-ion and sub-ion scales, i.e. at scales of the order of those involved in the electron temperature anisotropy generation process outlined above. On the other hand, the oblique whistler-cyclotron instability typically develops for high perpendicular electron temperature anisotropies. This is a usual and well-studied mechanism in the magnetosheath (among other environments) for generating waves \citep{Gary_etal_1996,Gary_etal_2005,Lee_etal_2018,Giagkiozis_etal_2018,Svenningsson_etal_2022,Svenningsson_etal_2024}. As we will show later, the strong electron temperature anisotropy produced by the large-scale electron velocity shears, is making the plasma unstable to the oblique whistler-cyclotron instability, causing the $\delta B_z$ fluctuations described previously.

\begin{figure}[ht]
\centering
\includegraphics[width=\linewidth]{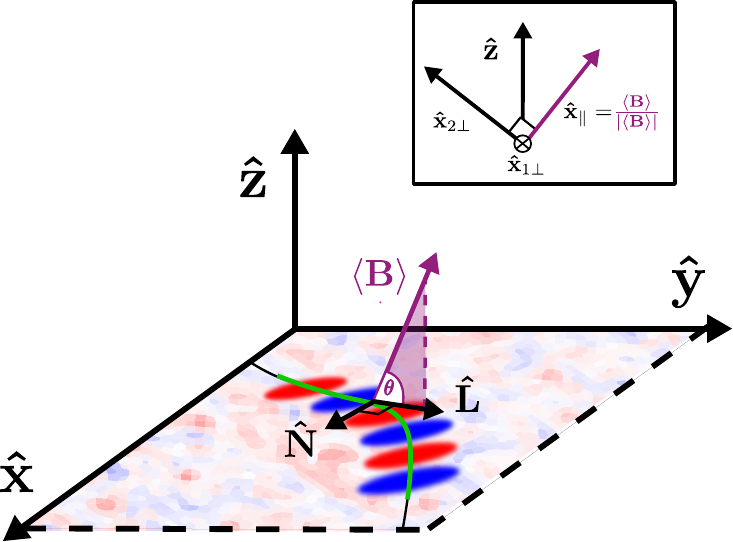}
\caption{Illustration of the coordinate systems defined in this work. The unit vectors $\hat{\mathbf{x}}$ and $\hat{\mathbf{y}}$ correspond to the in-plane directions while  $\hat{\mathbf{z}}$ is directed perpendicular to the plane. The orthogonal system of coordinates aligned with the low pass filtered magnetic field is given by $\hat{\mathbf{x}}_\parallel = \langle \mathbf{B} \rangle/|\langle \mathbf{B} \rangle|$, $\hat{\mathbf{x}}_{1\perp} = \hat{\mathbf{z}}\times \langle \mathbf{B}\rangle/\sqrt{\langle B_x\rangle^2+\langle B_y\rangle^2}$, and $\hat{\mathbf{x}}_{2\perp} = \langle \mathbf{B}\rangle \times \hat{\mathbf{x}}_{1\perp} /\left(|\langle \mathbf{B}\rangle |\sqrt{\langle B_x\rangle^2+\langle B_y\rangle^2}\right)$, with $\hat{\mathbf{x}}_\parallel = \hat{\mathbf{x}}_{1\perp}\times \hat{\mathbf{x}}_{2\perp}$. Notice that $\hat{\mathbf{x}}_{\parallel}$ is parallel to the background magnetic field direction, $\hat{\mathbf{x}}_{1\perp}$ is in-plane and perpendicular to the background magnetic field and $\hat{\mathbf{x}}_{2\perp}$ completes the orthogonal system of coordinates. The LNz orthogonal system of coordinates is aligned with the in-plane magnetic field and is given by $\hat{\mathbf{L}} = \left(\langle B_x \rangle \hat{\mathbf{x}} + \langle B_y \rangle \hat{\mathbf{y}}\right)/\sqrt{B_x^2 + B_y^2}$, $\hat{\mathbf{N}} = \hat{\mathbf{L}}\times\hat{\mathbf{z}}$ and $\hat{\mathbf{z}}$. $\hat{\mathbf{L}}$ and $\hat{\mathbf{N}}$ are denoted as "longitudinal" and "normal" directions, respectively. Notice that $\hat{\mathbf{L}}$ and $\hat{\mathbf{N}}$ are parallel and perpendicular to the in-plane background magnetic field, respectively.}
\label{fig:LNz_Coordinates}
\end{figure}

To characterize the magnetic field fluctuations, we can see in panels (a)-(e) of Figure \ref{fig:Whistler_characteristics} different plasma quantities calculated along the cuts highlighted in Figure \ref{fig:Anisotropy_instability} at $t=125\Omega_e^{-1}$ when the wave has already developed. In panel (a) of this figure, we show the different components of the magnetic field. The magnetic field points mainly in the $-x$ direction, with small fluctuations on the other components. Given the small amplitude of these fluctuations with respect to the background magnetic field, we will later interpret their formation using linear theory (see below). We use the background magnetic field (the low-pass-filtered magnetic field) to introduce a local coordinate system defined by the following unit vectors: $\hat{\mathbf{x}}_{1\perp} = \hat{\mathbf{z}}\times \langle \mathbf{B}\rangle/\sqrt{\langle B_x\rangle^2+\langle B_y\rangle^2}$, $\hat{\mathbf{x}}_{2\perp} = \langle \mathbf{B}\rangle \times \hat{\mathbf{x}}_{1\perp} /\left(|\langle \mathbf{B}\rangle |\sqrt{\langle B_x\rangle^2+\langle B_y\rangle^2}\right)$ and $\hat{\mathbf{x}}_\parallel = \langle \mathbf{B}\rangle/|\langle \mathbf{B}\rangle| = \hat{\mathbf{x}}_{1\perp}\times\hat{\mathbf{x}}_{2\perp}$. With these definitions, $\hat{\mathbf{x}}_{1\perp}$ lies in the x-y plane, while $\hat{\mathbf{x}}_{2\perp}$ has an out-of-plane component. In panel (b) of Figure \ref{fig:Whistler_characteristics}, we see the high-pass-filtered components of the magnetic field in the magnetic field-aligned system of reference defined above. We selected a sub-region of the cut, highlighted in yellow, to calculate the hodogram of magnetic field fluctuations in panel (g), with red and blue dots indicating the initial and final points of the trajectory, respectively. We see that the wave has almost right-handed circular polarization. Also, we show the Fourier power spectrum of the high-pass filtered magnetic fluctuations, in panel (f), revealing an intense spectral peak at $k d_e = 0.81$, i.e. at electron scales. From panel (c), we show the electron and ion densities along the cut. There is quasi-neutrality, with almost no density fluctuations, implying that these fluctuations are incompressible. In panels (d) and (e) we show the components of the electron and ion fluid velocities, respectively. In these panels, we used the subscripts $L$ and $N$ for "longitudinal" and "normal" coordinates, where the longitudinal direction $\hat{\mathbf{L}}$ is defined as parallel to the in-plane magnetic field, which is along the wave propagation direction. The normal direction is in-plane and perpendicular to the direction of wave propagation, defined as $\hat{\mathbf{N}}=\hat{\mathbf{L}}\times\hat{\mathbf{z}}$. From now on, we will refer to this system of reference as the "LNz" system of coordinates. A sketch of the LNz system of coordinates and the local coordinate system aligned with the background magnetic field is illustrated in Figure \ref{fig:LNz_Coordinates}. We see in panel (d) of Figure \ref{fig:Whistler_characteristics} that there are no significant fluctuations in the longitudinal component of the electron velocity $u_{eL}$, unlike the other components. This is also consistent with an incompressible wave. Furthermore, ions do not exhibit any fluctuations as it is shown in panel (e). We thus conclude that only the electrons are coupled to the wave. All these properties are consistent with an oblique whistler wave \citep{Krall_Trivelpiece_1986, Gary_1993}.

To further support our interpretation and compare simulation data with theoretical predictions, we use the DIS-K linear solver \citep{Lopez_etal_2021,Lopez_etal_2021_b} to analyze the stability of the plasma. The parameters used to initialize the linear solver have been calculated as averages over the cuts marked in the panels of the second column of Figure \ref{fig:Anisotropy_instability}, at $t=75\Omega_e^{-1}$, when the wave has just developed. The input values used in the linear solver are given in Table \ref{tab:LinearSolverData}. Panels (h)-(j) of Figure \ref{fig:Whistler_characteristics} show the shaded isocontours for several quantities calculated via the linear solver, i.e. the real part $\omega$ and the imaginary part $\gamma$ (the growth/damping rate) of the wave frequency, and the electric field polarization $P=\text{Re}\{iE_{\perp 1}/E_{\perp 2}\}$, as functions of the normalized wavenumber $k d_e$, and of the angle between the wave propagation direction and the background magnetic field $\theta$. In all these panels, we have marked the wavenumber deduced from the power spectra analysis with a vertical dashed black line, $k d_e = 0.81$. These quantities are calculated for angles $\theta$ between 18 and 30, which is in the range for where the wave has been propagating, as we can see in panels (h) and (i) of Figure \ref{fig:Anisotropy_instability}. In panel (h) of Figure \ref{fig:Whistler_characteristics}, we see that for the wavenumber deduced from the simulation, the linear solver predicts a frequency of $\omega/\Omega_e \sim 0.42$, which is in the frequency range where waves can be found. Also, in panel (i) of the exact figure, it is shown that for the considered wavenumber $k d_e= 0.81$, there is a positive maximum in the imaginary part of the wave frequency $\gamma$. Therefore, the linear solver predicts an instability with a growth rate $\gamma$ that decreases with increasing $\theta$. As it is shown in the contour lines of panel (i), the growth rate is in the range $\gamma/\Omega_e \in [0.01,0.025]$, which implies a characteristic time for the growth of the wave of $T\Omega_e \in [40,100]$, in the same range as the growth time observed in the simulation. This supports our claim that the magnetic field perturbations are associated with a wave produced by a temperature anisotropy instability. Panel (j) shows that the linear solver predicts an almost right-handed circular polarization for the electric field (which has the same polarization as the magnetic field due to the Faraday law $\nabla \times \mathbf{E} = (-1/c)\partial\mathbf{B}/\partial t$, that matches with what we observe in the hodogram in panel (g).

Hence, we conclude that the linear solver predictions are consistent with the properties of the wave observed in the simulation. Furthermore, the results from both the simulation and the linear theory show a right-handed and unstable wave at electron scales. Therefore, the direct analysis of simulation data, combined with results from the linear solver, indicates that the observed electron scale fluctuations are consistent with the propagation of an oblique whistler wave, generated by an oblique whistler-cyclotron instability due to the electron temperature anisotropy. 

\renewcommand{\tabcolsep}{20pt}
\begin{table}
\centering
\caption{Parameters considered for the DIS-K linear solver taken from the mean values of the 1D cut crossings for $t=75\Omega_e^{-1}$ and showed in the second column of Figure \ref{fig:Anisotropy_instability}.}
\label{tab:LinearSolverData}
\begin{tabular}{cc}
\toprule
Plasma parameters & Mean values\\ 
\midrule
$m_i/m_e$ & 100 \\
$\omega_{pe}/|\Omega_e|$ & 4.1 \\
$A_e$ & 2.1 \\
$A_i$ & 1.8 \\
$\beta_{\parallel e}$ & 0.3 \\
$\beta_{\parallel i}$ & 1.8 \\
\bottomrule
\end{tabular}
\end{table}	

\subsection{Whistler-to-Bernstein
mode conversion and electron scale MH formation} 

\begin{figure*}[ht]
\centering
\includegraphics[width=\linewidth]{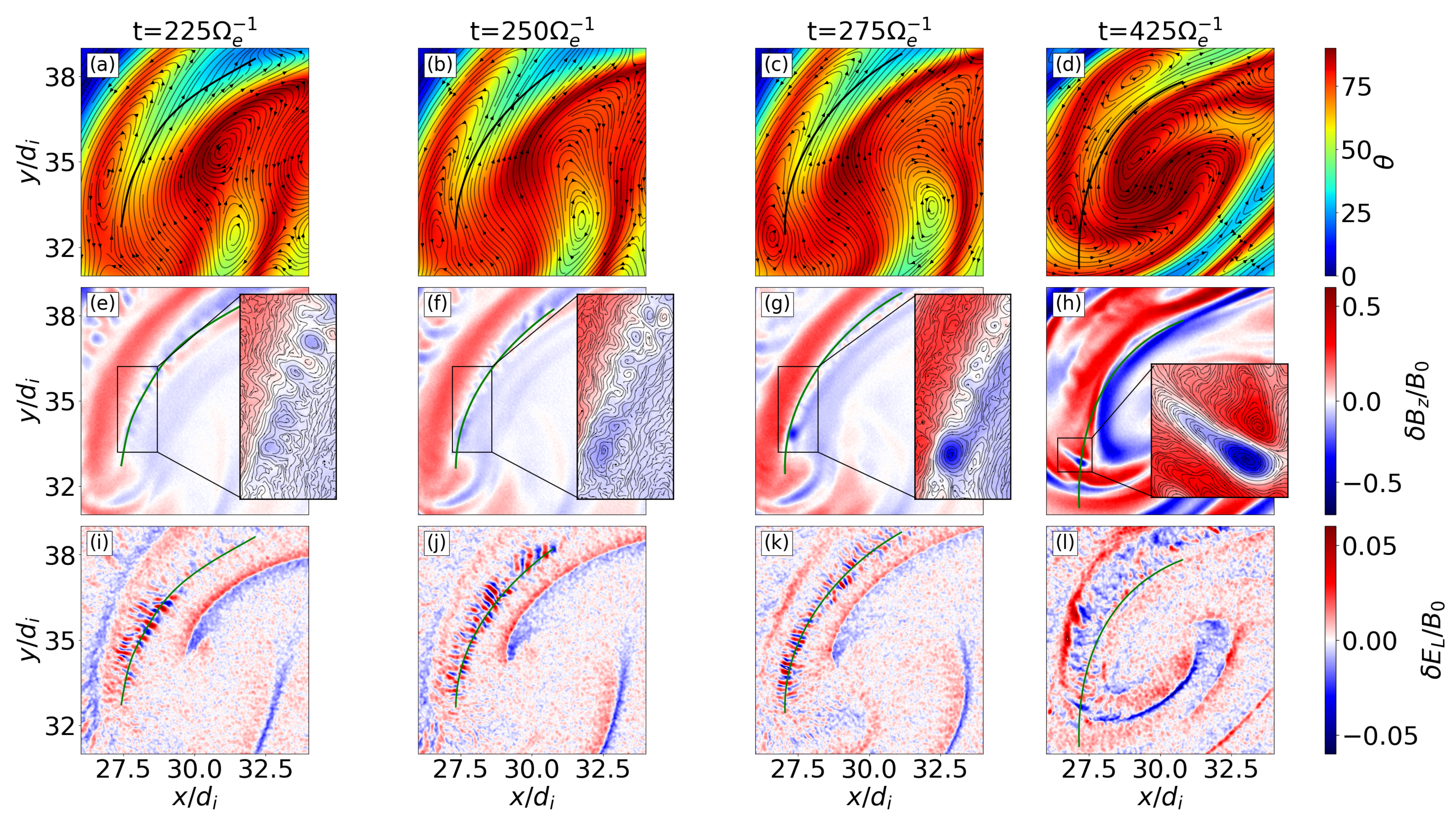}
    \caption{Panels (a)-(d): shaded isocontours of the angle $\theta$ between the low pass filtered magnetic field and the plane with the black streamlines representing the in-plane low-pass filtered total current density $\langle \mathbf{J} \rangle = \langle\mathbf{J}_i\rangle + \langle\mathbf{J}_e\rangle$. Panels (e)-(h):  shaded isocontours of the high pass filter of the out-of-plane component of the magnetic field $\delta B_z$ with a zoom on the electron scale vortices, with the black streamlines representing the in-plane high pass filtered electron current density $\delta \mathbf{J}_e$. Panels (i)-(l): shaded isocontours of the longitudinal component of the high-pass filtered electric field $\delta E_L$. The panels are divided into four columns showing the time sequence in the following times $t=225\Omega_e^{-1}$, $t=250\Omega_e^{-1}$, $t=275\Omega_e^{-1}$, and $t=425\Omega_e^{-1}$. The lines marked on times $t=225\Omega_e^{-1}$, $t=275\Omega_e^{-1}$, and $t=425\Omega_e^{-1}$ represent the 1D cut crossings used to show the wave characteristics for the longitudinal waves (see below). }
\label{fig:Bernstein_and_VortexMerging}
\end{figure*}

\begin{figure*}[ht]
\centering
\includegraphics[width=0.9\linewidth]{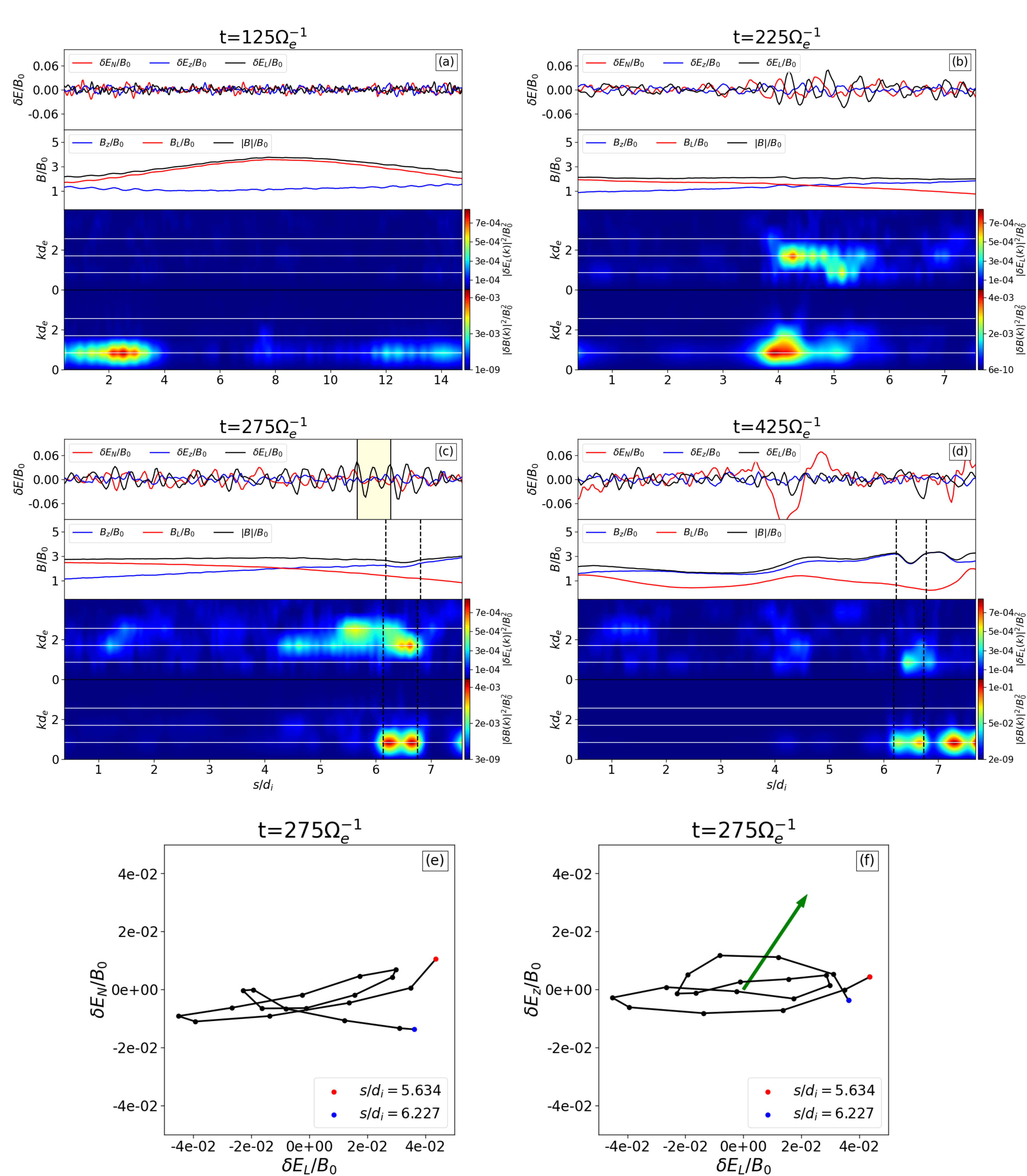}
    \caption{Panels (a)-(d): each panel shows the components of the high pass electric field $\delta E_L$ in LNZ coordinates, the total magnetic field components and the longitudinal electric field and total magnetic field spectrogram for times $t=125\Omega_e^{-1}$, $t=225\Omega_e^{-1}$, $t=275\Omega_e^{-1}$ and $t=425\Omega_e^{-1}$ respectively over the 1D cut crossings showed in figures \ref{fig:Anisotropy_instability} and \ref{fig:Bernstein_and_VortexMerging}. The black-shaded vertical lines in panels (c) and (d) delimit the region of the magnetic field depression, which later develops as the electron scale MH. The three white horizontal lines on the spectrograms represent the harmonics of the frequency i.e. $kd_e = 0.81, 1.62, 2.43 $. Panels (e)-(f): hodograms of the high-pass filtered components of the electric field in LNZ coordinates over the region highlighted in yellow in panel (c) for $t=275\Omega_e^{-1}$. The green arrow represents the direction of the mean background magnetic field.}
\label{fig:Bernstein_characteristics}
\end{figure*}

\begin{figure*}[ht]
\centering
\includegraphics[width=\linewidth]{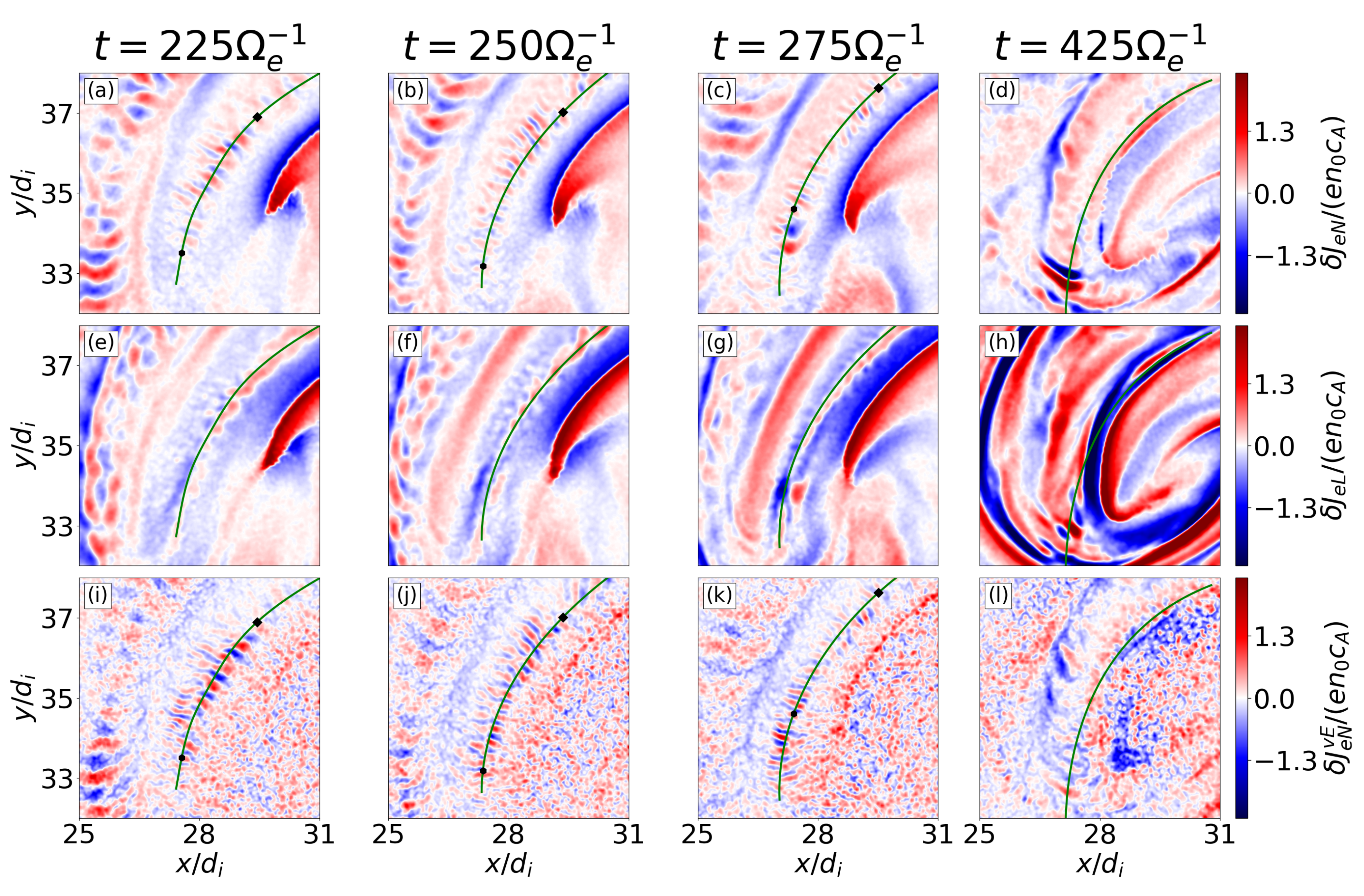} 
    \caption{Panels (a)-(d): shaded isocontours of the normal component of the high-pass filtered electron current density $\delta J_{eN}$. Panels (e)-(h): shaded isocontours of the longitudinal component of the high-pass filtered electron current density $\delta J_{eL}$. Panels (i)-(l): shaded isocontours of the normal component of the electron velocity drift $\delta J_{eN}^{vE} = \left(-e \langle n_e\rangle \delta \mathbf{E}\times \langle \mathbf{B}\rangle/|\langle \mathbf{B}\rangle|^2\right)\cdot \hat{\mathbf{N}}$ due to the high pass filtered longitudinal electric field and the low pass filtered magnetic field. The panels are divided into four columns showing the time sequence in the following times $t=225\Omega_e^{-1}$, $t=250\Omega_e^{-1}$, $t=275\Omega_e^{-1}$, and $t=425\Omega_e^{-1}$. The 1D cut crossings represented by the green lines are the same as in earlier figures for their respective times. The black points delimit the region used for the correlation between the electron current density observed in the simulation and the one due to the electron drift velocity (see below).}
\label{fig:Currents}
\end{figure*}

\begin{figure*}[ht]
\centering
\includegraphics[width=\linewidth]{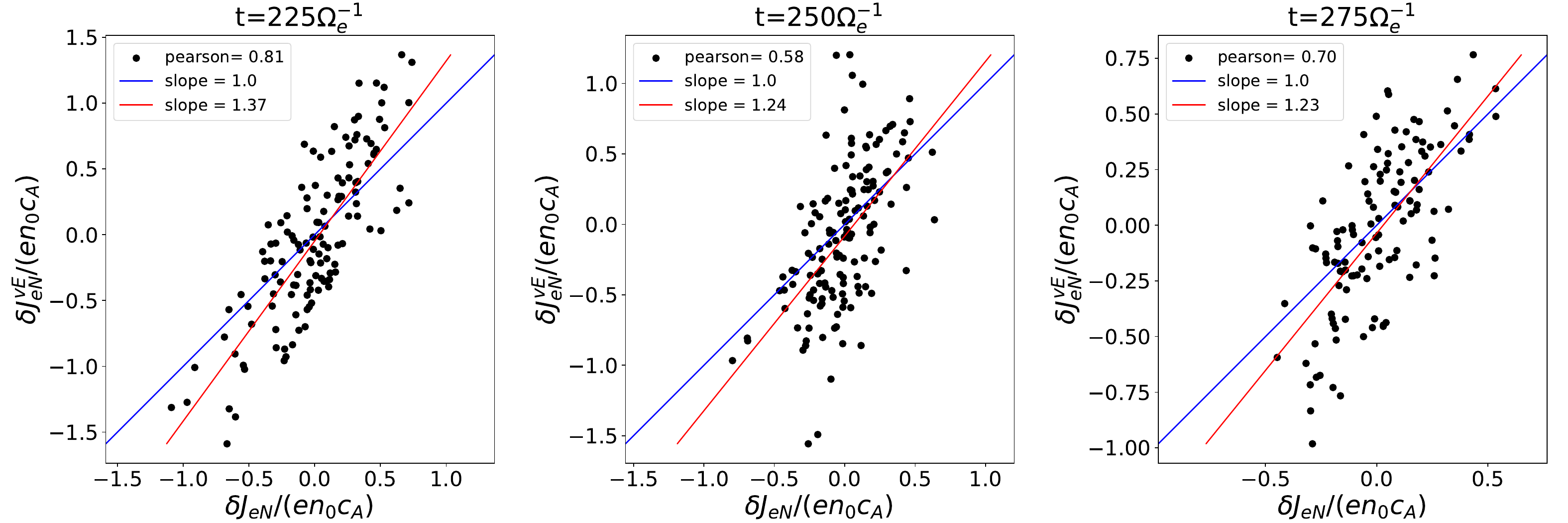} 
    \caption{Correlation between the normal components of the high pass filtered electron current density $\delta J_{eN}$ and the drift electron current density $\delta J_{eN}^{vE}$ for the 1D cut crossings delimited by black points on figure \ref{fig:Currents} for times $t=225\Omega_e^{-1}$,$t=250\Omega_e^{-1}$ and $t=275\Omega_e^{-1}$. The blue lines represent the straight line with slope 1, and the red lines represent the straight line calculated using the least squares method.}
\label{fig:DriftCurrent_correlation}
\end{figure*}

In this subsection, we will analyze the subsequent development of the wave in its nonlinear stage, and we will show how these fluctuations are correlated with the formation of vortices that give rise to the electron scale MH.

In Figure \ref{fig:Bernstein_and_VortexMerging} we show at different times the shaded isocontours of the angle $\theta$ between the magnetic field and the plane (with the low-pass filtered total current density $\langle \mathbf{J}\rangle = \langle \mathbf{J}_e\rangle + \langle\mathbf{J}_i\rangle$ streamlines represented by black lines), the high-pass-filtered magnetic field in the out-of-plane direction $\delta B_z$, and the longitudinal component of the high-pass-filtered electric field $\delta E_L$. These figures describe the wave evolution and the formation of the electron scale MH over time. We will describe this process step by step in the following.

At time $t=225\Omega_e^{-1}$, we can see in panel (a) of Figure \ref{fig:Bernstein_and_VortexMerging} that there is a large-scale current density configuration (black lines) which forms a large-scale vortex that is responsible for increasing the out-of-plane magnetic field, and thus the angle $\theta$ between the magnetic field and the plane. The wave then passes from a region with an angle of propagation mostly parallel to the background magnetic field ($\theta$ less than 25 degrees) to a highly oblique propagation, with $\theta$ higher than 50 degrees. We can see in panel (e) that at this stage the wavefront of magnetic field fluctuations is no longer V-shaped, as it was in panels (b) and (c) of Figure \ref{fig:Anisotropy_instability}. Fluctuations now resemble a chain of "bubbles", meaning that the wave has evolved into a different structure. In panel (i), at time $t=225\Omega_e^{-1}$, significant longitudinal fluctuations in the electric field appear. At later times, in panels (j) and (k), we see that the longitudinal fluctuations of the electric field are still propagating. In the same region where there are significant values of $\delta E_L$ at $t$ between $225\Omega_e^{-1}$ and $275\Omega_e^{-1}$, we see in panels (e)-(g) the presence of bubble-shaped magnetic field fluctuations. These fluctuations produce a magnetic field dip, which will later develop into the MH, as discussed in section 3.1. At the time $t=425\Omega_e^{-1}$ when this magnetic field dip is well developed (as shown in panel (h)), we no longer see significant fluctuations in the longitudinal component of the electric field, as shown in panel (l).

In panels (e)-(h) of Figure \ref{fig:Bernstein_and_VortexMerging} we see a zoom into the bubble-shaped magnetic field fluctuations, with the high-pass-filtered electron current density streamlines represented by black lines. It is clear from these panels that these "bubbles" are vortices produced by electron ring currents. In panel (f), at time $t=250\Omega_e^{-1}$, we can see that two of these vortices are merging, generating a larger vortex with a deeper decrease in the out-of-plane magnetic field. The large vortex subsequent evolution is shown in panel (g), at $t=275\Omega_e^{-1}$. Finally, this vortex sustained by the electron current ring is convected by the plasma and evolves until it becomes a stable coherent structure, as we will explain below. 

Vortex merging is a typical phenomenon well studied in hydrodynamics via both experimental, numerical, and analytical studies \citep{Melander_etal_1988, CERRETELLI_and_WILLIAMSON_2003}. In the hydrodynamic case, the merging depends highly on the initial conditions \citep{MEUNIER_etal_2005, Leweke_etal_2016}. For plasmas, vortex merging has been mainly studied in the context of Kelvin-Helmholtz instabilities, using two-fluid models \citep{Akira_1997, Tenerani_etal_2011}, and in dusty plasmas \citep{Dharodi_etal_2024}. In the case of plasmas, the vortex dynamics are more complex due to the presence of the magnetic field generated by the vortical currents and the different behaviors of electrons and ions \citep{Tur_and_Yanovsky_2017}. A detailed investigation of the dynamics responsible for merging the indicated vortices goes beyond the scope of this paper. Here, we aim to point out that vortex merging produces a larger and more intense current ring that further locally reduces the magnetic field amplitude. This larger vortex then survives and develops into an electron-scale MH, whose properties have been described in section 3.1, while the other vortices are destroyed by the background turbulent dynamics.

In the following, we will analyze in more detail the characteristics of the fluctuating longitudinal electric field discussed above. This will allow us to show in the next subsection that these fluctuations are responsible for the formation of the chain of vortices shown in Figure \ref{fig:Bernstein_and_VortexMerging}.
Using data from the 1D cuts marked by colored lines in Figures \ref{fig:Anisotropy_instability} (for $t=125\Omega_e^{-1}$) and \ref{fig:Bernstein_and_VortexMerging} (for $t\Omega_e \in \{225,275,425\}$), we show in Figure \ref{fig:Bernstein_characteristics}, for different times, the components of the high-pass-filtered electric field and the total magnetic field in LNz coordinates, together with spectrograms of the longitudinal component of the electric field $\delta E_L$ and of the magnetic field magnitude. On top of the spectrograms, we show three white horizontal lines representing the fundamental, second, and third harmonics of the wave discussed before, i.e. $k d_e = 0.81, 1.62, 2.43$. Panel (a) shows results at $t=125\Omega_e^{-1}$ when the wave is in its linear stage, and we see that there are no significant longitudinal electric field fluctuations, as observed in the spectrograms. We also see that the magnetic field fluctuations exhibit a peak at the wavenumber deduced above via the FFT analysis over the 1D crossing. Later, at $t=225\Omega_e^{-1}$, panel (b), we see that significant longitudinal electric field fluctuations develop, consistent with what we have shown previously in panel (i) of Figure \ref{fig:Bernstein_and_VortexMerging}. The spectrogram shows that these longitudinal electric field fluctuations have peaks in correspondence with the white lines, i.e. they match the harmonics of the wave. At the time $t=225 \Omega_e^{-1}$ the peak frequency of $\delta E_L$ matches with the first and second harmonics, then at $t=275\Omega_e^{-1}$ the third harmonic is excited as well. At both times $t=225\Omega_e^{-1}$ and $t=275\Omega_e^{-1}$, the second harmonic carries more energy than the first and third harmonics. Then, the spectrogram shows a correlation between the longitudinal electric field fluctuations and the magnetic field fluctuations, with the first having mostly half of the wavelength of the second. As longitudinal electric field fluctuations develop, the wave changes its nature. As shown above, at this stage, fluctuations have turned into electron vortices, producing alternating magnetic field dips and peaks. Both the peaks and the dips have the same length of the longitudinal electric field fluctuations excited at the second harmonic of the wave. In panel (c), at time $t=275\Omega_e^{-1}$, we see that the strong longitudinal electric field fluctuations are correlated with the region where the two vortices have merged, forming the magnetic dip. The location of this magnetic dip is marked by two dashed vertical black lines in panels (c) and (d). Just before the magnetic dip location, we highlighted in yellow a region that covers two wavelengths of the electrostatic wave. For this region, we show in panel (e) the hodogram of the normal and longitudinal components of the electric field, and in panel (f) a hodogram of the out-of-plane and longitudinal components. We see that the polarization is almost linear, with small fluctuations in the plane perpendicular to the direction of propagation of the wave. Moreover, panel (f) shows the projection of the mean magnetic field (over the yellow region) in the z-L plane, indicated by a green arrow, which is at an angle $\theta \sim 56$ with respect to the longitudinal direction, mainly aligned in the out-of-plane $z$ direction. Thus, these fluctuations correspond to quasi-electrostatic modes, with properties consistent with oblique Bernstein waves \citep{Krall_Trivelpiece_1986}. Therefore, hereafter we will refer to them as quasi-electrostatic Bernstein-like modes. The development of these modes can be explained in the following way. At $t=125\Omega_e^{-1}$, the direction of wave propagation (i.e. the longitudinal direction) is almost aligned with the direction of the magnetic field, as seen in panel (a) of Figure \ref{fig:Bernstein_characteristics} where the component $B_L$ almost coincides with the total magnitude of the magnetic field. As the wave propagates through the inhomogeneous plasma, it enters a region where the out-of-plane component of the magnetic field increases. Therefore, the angle between the wave's direction and the magnetic field's orientation increases. Hence, the propagation goes from being quasi-parallel to quasi-perpendicular to the magnetic field. In other words, the propagation direction of the wave becomes highly oblique with respect to the background magnetic field, as the wave travels through the inhomogeneous environment. Now, whistler waves do not propagate perpendicular to the magnetic field, so as the magnetic field becomes more aligned to the out-of-plane direction, the component of this fluctuation is suppressed. On the other hand, Bernstein mode propagates perpendicular to the magnetic field, so as $B_z$ increases, perpendicular electric field perturbations develop as a consequence of the excitation of Bernstein modes, and the fluctuation decays from a quasi-parallel wave to a quasi-perpendicular Bernstein mode. Indeed, in panel (b) of Figure \ref{fig:Bernstein_characteristics}, we see that the Bernstein-like mode appears when $B_L$ overcomes $B_z$, i.e., when the angle between the wave propagation and the background magnetic field is higher than $45$ degrees. At the time $t=275\Omega_e^{-1}$, we see in panel (c) of the same figure that $B_z$ further increases, and the Bernstein mode fluctuations propagate in a more oblique direction to the magnetic field. The same effect is seen in Figure \ref{fig:Bernstein_and_VortexMerging}, where the electrostatic fluctuations are more prominent in high-angle regions, and as time increases, the electrostatic fluctuations extend further in space. One theoretical argument to interpret the whistler-to-Bernstein mode conversion is considering a linear expansion of kinetic plasma equations for whistler wave propagation, following the approach described in \citep{YuX2021}. When taking this expansion to orders higher than one, it is possible to obtain a forced wave equation where a nonlinear current sustained by the whistler waves acts as an antenna that induces the generation of higher harmonics, depending on the order of the linear expansion. Nevertheless, since the background plasma we have in the simulation is highly inhomogeneous and turbulent, the plasma response, described by the nonlinear conductivity tensor, will change in space as the whistler fluctuation propagates, so the original whistler wave will excite a different mode, in this case, a quasi-electrostatic Bernstein-like fluctuation. The mode conversion of waves into quasi-electrostatic nonlinear harmonics has also been observed by MMS in the Earth's magnetosheath, as reported in \citet{Xu_etal_2024}. Afterward, the longitudinal electric field component of the Bernstein-like mode couples to the plasma, producing the series of vortices that finally merge into a single EVMH, as discussed above. The dynamics responsible for forming the chain of vortices are discussed in the next section.

\subsection{Vortex formation due to currents induced by the quasi-electrostatic Bernstein-like mode}

In this subsection, we will discuss the mechanism for the formation of the vortices and their relation with the longitudinal electric field fluctuations. Also, we will discuss how the electron ring current that sustains the magnetic dip evolves to the point discussed in section 3.1, producing the fully developed electron scale MH.  

\begin{figure}[ht]
\centering
\includegraphics[width=\linewidth]{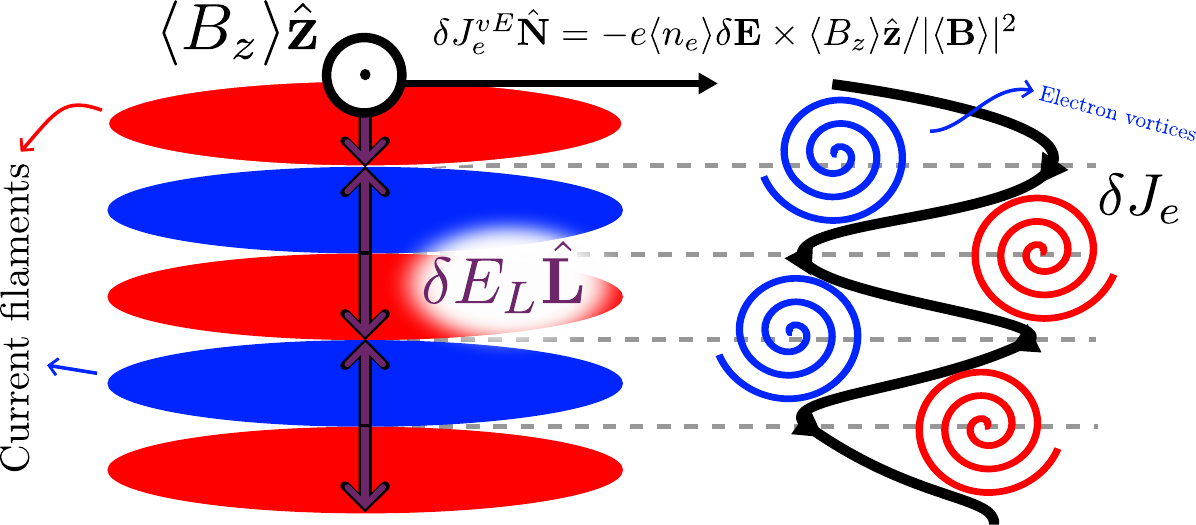}
\caption{Illustration of the mechanism presented in this work for the formation of the chain of electron scale vortices observed in the simulation. The longitudinal electric field fluctuations of the Bernstein-like modes, when interacting with the background magnetic field via the $\mathbf{E}\times\mathbf{B}$ drift, produce current filaments which oscillate in space. These current filaments oriented in the normal direction $\hat{\mathbf{N}}$ are represented by colored ellipses, with red filaments pointing to the right and blue filaments pointing to the left. The combination of these current filaments and the longitudinal background electron current produces a "zig-zag" small-scale electron current that produces the chain of electron scale vortices seen in the simulation. The blue and red vortices represent a dip and a hump of the magnetic field magnitude respectively.}
\label{fig:vortices_ilustration}
\end{figure}

Figure \ref{fig:Currents} shows the shaded isocontours of the normal component of the electron $\textbf{E}\times\textbf{B}$ drift due to the electric field fluctuations $\delta \mathbf{J}_{e}^{vE} = -e \langle n_e\rangle \delta \mathbf{E}\times \langle \mathbf{B}\rangle/|\langle \mathbf{B}\rangle|^2$, along with the normal $\delta J_{eN}$ and the longitudinal $\delta J_{eL}$ components of the high-pass-filtered electron current density. Different times are represented, following the evolution of these current components in the region where the MH develops. In this figure, the 1D cuts marked in green are the same as those analyzed above in Figure \ref{fig:Bernstein_and_VortexMerging}. In panels (a)-(d) we see that in the same region where the quasi-electrostatic Bernstein-like mode appears, at $t\Omega_e \in \{225,250,275\}$, $\delta J_{eN}$ shows the presence of a chain of current filaments, perpendicular to the direction of propagation of the wave. In panels (e)-(h) we can see that this sequence of normal electron current filaments is accompanied by two regions of longitudinal electron current density with opposite directions. The combination of normal and longitudinal currents produces the zigzag-shaped current streamlines shown in panels (e)-(h) of Figure \ref{fig:Bernstein_and_VortexMerging}, which sustain the chain of vortices associated with magnetic peaks and dips. At time $t=275\Omega_e^{-1}$, we see in panels (c) and (g) of Figure \ref{fig:Currents} that in the region next to the lowest (from top to bottom) black dot over the 1D cut, vortices are merged. This is indicated by the strong bipolar normal and longitudinal currents, highlighting the presence of a vortex of size larger than those in the chain at previous times. On the other hand, in panels (i)-(l) we show the normal component of the current density produced by the $\textbf{E}\times\textbf{B}$ drift due to longitudinal electric field fluctuations, i.e. $\delta J_{eN}^{vE}$. We see a marked correlation between $\delta J_{eN}^{vE}$ and $\delta J_{eN}$ at times $t\Omega_e^{-1} \in \{225,250,275\}$. However, by comparing panels (c) and (k), we see that in the region where the vortices have merged, $\delta J_{eN}^{vE}$ no longer matches $\delta J_{eN}$. In Figure \ref{fig:DriftCurrent_correlation} we show the scatter plots of $\delta J_{eN}$ vs $\delta J_{eN}^{vE}$, taken from the 1D cuts of Figure \ref{fig:Currents}, sampling these currents over the segments between the black diamonds and the black dots (indicated in Figure \ref{fig:Currents}), at times $t\Omega_e \in \{225,250,275\}$. These segments correspond to regions where $\delta J_{eN}^{vE}$ is stronger, without crossing the merged vortex area. We calculated the Pearson correlation coefficient between $\delta J_{eN}^{vE}$ and $\delta J_{eN}$ for each scatter plot, which is equal to 0.81 at $t=225\Omega_e^{-1}$, 0.58 at $t=250\Omega_e^{-1}$ and $0.70$ at $t=275\Omega_e^{-1}$, showing a good correlation between the two current components. In Figure \ref{fig:DriftCurrent_correlation}, the straight blue lines represent the relation $\delta J_{eN}^{vE}\!=\!\delta J_{eN}$, while red lines are calculated by fitting the distributions of data points with a straight line, with their corresponding slope indicated in the label of the figure. We see that $|\delta J_{eN}^{vE}|$ tends to be larger than $|\delta J_{eN}|$ (between $23\%$ and $37\%$ times larger). This mismatch arises from the fact that the total current contains other contributions besides the $\textbf{E}\times\textbf{B}$ component, e.g., contributions coming from pressure gradient drifts, magnetic field drifts, and kinetic effects. Moreover, numerical noise in the electric field may also affect the correlation.

Hence, the analysis outlined above suggests that the longitudinal electric field fluctuations of the Bernstein-like mode, combined with the local magnetic field, produce filamentary $\textbf{E}\times\textbf{B}$ drift currents that are responsible for the development of the chain of electron scale vortices as it is illustrated in Figure \ref{fig:vortices_ilustration}. As discussed in previous sections, this sequence of vortices finally merges into a bigger vortex that reduces the local magnetic field, ultimately evolving into the fully developed electron scale MH shown and discussed in section 3.1. It is important to note that, due to vortex merging, the size of the magnetic hole is twice the wavelength of the waves that drive its formation. Therefore, the scale of the magnetic hole depends not only on the wavelength of the fluctuations that generate it, but also on the competition between vortex merging and the background turbulence dynamics, where the latter tends to destroy the magnetic depressions and peaks produced by the wave, as shown in Figure \ref{fig:Bernstein_and_VortexMerging}.

\section{Conclusions}

In this work, we have studied the self-consistent formation and the properties of an electron-scale MH observed in a fully kinetic 2D simulation of plasma turbulence, initialized with parameters consistent with those observed in the Earth's magnetosheath.

We have identified and characterized turbulent-driven dynamics capable of generating MHs at scales of the order of a few electron inertial lengths. Step by step, we have dissected and analyzed this particular generation dynamic, highlighting how large-scale fluctuations transfer their energy to small-scale fluctuations, setting up the conditions for forming electron-scale MHs. The dynamics responsible for the generation of the specific MH we analyzed consist of several steps, summarized as follows. 1: Large-scale turbulent velocity shears produce localized regions with strong perpendicular electron temperature anisotropy. 2: These regions quickly become unstable, producing quasi-parallel whistler waves. 3: As waves propagate over the quickly varying and inhomogeneous turbulent background, they develop a quasi-electrostatic component, evolving into Bernstein-like modes. 4: The electrostatic fluctuations of Bernstein-like modes induce filamentary $\textbf{E}\times\textbf{B}$ drift currents that turn the wave into a train of current vortices. 5: These vortices finally merge into a larger vortex that reduces the local magnetic field magnitude, ultimately evolving into a coherent electron scale MH.

The properties of the fully developed electron scale MH observed in the simulation are consistent with typical measurements of small-scale MHs in the Earth's magnetosheath \citep{Shi_etal_2024}. In particular, we have shown that the electron scale MH is characterized by an electron current ring that sustains the structure. We have shown that the electron scale MH hosts hot electrons with large pitch angles, which is associated with a sharp increase in density and the perpendicular electron temperature inside the structure. On the other hand, ions are not coupled to the MH due to its small size, and ion quantities do not show any variations correlated with the magnetic field magnitude depression. These features have also been observed in other turbulent sub-ion scale MHs in simulations \citep{Haynes_etal_2015,Roytershteyn_etal_2015,Arro_etal_2023}. We have also shown that the electron scale MH has nontrivial kinetic properties. The EVDF inside the MH has a "mushroom" shape, with two populations, a hot anisotropic ring-shaped population and a colder and less anisotropic population, aligned with the magnetic field's parallel direction. 

This analysis not only highlights the important role of velocity shears in the development of the turbulence and the formation of sub-ion scale MHs, as discussed also in \citet{Arro_etal_2023}, but it also shows the importance of nonlinear wave processes and cross-scale interactions in the generation of electron scale coherent structures. Other studies and observations have also suggested a correlation between wave propagation and the formation of MHs by nonlinear mechanisms. However, these studies mainly focus on large-scale MHs \citep{Tsurutani_etal_2002_b, Xu_etal_2024}. On the other hand, in the case of small-scale MHs described by EMHD solitons, even if these are consistent with some observations in the plasma sheet \citep{Ji_etal_2014, Li_etal_2016}, they cannot explain non-propagating MHs (in the plasma rest frame), frequently observed in the Earth's magnetosheath \citep{Shi_etal_2024}. Therefore, we have provided new insights into the role of electron scale waves and nonlinear processes in the formation of electron scale MHs, and how the turbulence dynamics can drive such mechanisms. This study has potential applications to observations in the Earth's magnetosheath and other space turbulent environments such as the solar wind and the terrestrial magnetotail. The mechanisms we discuss also represent a way for the plasma to transfer energy from large scales to electron scales in turbulent scenarios, showing the important role of electron scale waves in turbulence dynamics. Recent numerical works have shown that turbulence is capable of generating MHs at large and near-ion scales, as discussed in \citet{Arro_etal_2023, Arro_2024}. Our results are complementary to these works, showing that turbulence can also produce MHs at electron scales. Although both in \citet{Arro_etal_2023} and the present work, electron velocity shears driven by the background turbulence dynamics give rise to the formation of vortices that stabilize as sub-ion scale MHs, the two mechanisms exhibit significant differences. In \citet{Arro_etal_2023}, thin an elongated electron velocity shears become unstable to the electron Kelvin-Helmholtz instability, breaking apart into electron vortices that eventually become sub-ion scale MHs. Instead, in the present article, the shear gives rise to a whistler wave that, in the end, is responsible for the formation of the vortices, which merge into an MH. Besides, in \citet{Arro_etal_2023} the resulting MHs are slightly larger, of the order of $\sim d_i$, while in the present study the MH has a size of $\sim 5d_e$, at electron scales.

The electron-scale MH structures potentially play a relevant role in the turbulent dynamics of the Earth's magnetosheath. Indeed, MHs have been linked to the occurrence of magnetic reconnection processes \citep{Zhong_etal_2019, Li_and_Zhang_2023}, it has been shown that they can generate waves thanks to their high-temperature anisotropy \citep{Smith_etal_1969, Huang_etal_2018, Yao_etal_2019}, and they can also accelerate, trap and scatter electrons, eventually heating the plasma \citep{Shi_etal_2024}. Concerning \citet{Yao_etal_2019} work, which shows how MHs can trigger electromagnetic and electrostatic fluctuations, we have demonstrated an opposite mechanism: waves drive the formation of MHs. It is important to highlight the differences between these mechanisms to clarify which of the two is at play. While in \citet{Yao_etal_2019} it is shown by a statistical study that most of the waves are localized inside MHs, in this study, the wave evolves before the development of the MH and is not spatially confined inside the structure, as the wave occupies a significantly larger portion of space. These properties highlight the different nature of the two phenomena mentioned above. 

In the present work, we have studied a particular chain of turbulence-driven mechanisms that lead to the formation of electron-scale MHs. Still, we do not claim that this is the only possible dynamic responsible for generating electron-scale MHs. It is worth mentioning that an important limitation of this work is the 2D geometry of our simulation, which does not allow us to explore the 3D structure of MHs. In \citet{Roytershteyn_etal_2015},
it has been shown, using 3D fully kinetic simulations, that MHs can develop self-consistently in turbulence, exhibiting a 3D cylindrical structure. It is still unclear which processes contribute to determining this specific 3D geometry. On the other hand, a 3D geometry could affect the mode conversion of waves into Bernstein waves, as wave propagation would not be confined to a 2D plane. Therefore, this 2D simulation represents a starting point for studying electron scale MHs and their generation in turbulent environments. More realistic 3D simulations will be considered in future works. 

The turbulent-driven mechanism responsible for the formation of electron-scale MHs also leads to a chain of magnetic field humps, sustained by electron vortices, which are then destroyed by the background turbulence dynamics, as is shown in Figure \ref{fig:Bernstein_and_VortexMerging}. These magnetic hump vortices, although short-lived, resemble the electron-scale magnetic peaks observed in \citet{Yao_etal_2018} using MMS data in the Earth's magnetosheath. These coherent structures have been detected on a scale of a few electron gyroradii, and an electron vortex is observed perpendicular to the field line. In future works, we will investigate whether under certain conditions these magnetic humps could stabilize before being annihilated by turbulence, leading to the formation of magnetic peaks like those observed in \citet{Yao_etal_2018}.

Electron-scale MHs have also been detected in the solar wind, the terrestrial magnetotail, planets, and cometary environments \citep{Russell_etal_1987, Huang_etal_2019, Yu_etal_2021, Chen_etal_2022}. In future works, we plan to study the role of turbulence in producing electron scale MHs via fully kinetic simulation initialized with parameters typical of the solar wind, the terrestrial magnetotail, comets, and the magnetosphere of other planets. These studies will help us to understand the details of electron scale MHs generation mechanisms in the aforementioned environments.

\begin{acknowledgements}

J.E.-T. acknowledges the support of ANID, Chile, through
National Doctoral Scholarship No. 21231291. P.S.M. acknowledges the support of ANID, Chile, through
the Fondecyt Grant No. 1240281. G. A. acknowledges support by the Laboratory Directed Research and Development program of Los Alamos National Laboratory under project number 20258088CT-SES. This research was supported by the International Space Science Institute (ISSI) in Bern, through the ISSI International Team project 24-627: Magnetosheath structures as seen by spacecraft observations and numerical simulations, and project 24-612: Excitation and Dissipation of Kinetic-Scale Fluctuations in Space Plasmas. The authors gratefully acknowledge the Gauss Centre for Supercomputing (GCS) e.V. (www.gauss-center.eu) for funding this project by providing computing time on the GCS Supercomputer SuperMUC-NG at Leibniz Supercomputing Centre (www.lrz.de).

\end{acknowledgements}

%





\bibliography{References}{}
\bibliographystyle{aasjournal}



\end{document}